\documentclass[12pt]{iopart}

\usepackage{iopams} 

\usepackage{epsfig,latexsym,amssymb,cite}
\usepackage{graphicx}
\usepackage{bm}

\newcommand{\as}{\alpha_s}
 \def\eq#1{{Eq.~(\ref{#1})}}
\def\fig#1{{Fig.~\ref{#1}}}
\newcommand{\msbar}{\mu_{\overline {MS}}}
\def\peq#1{{(\ref{#1})}}
\newcommand{\amu}{\alpha_\mu}

\begin{document}

\title{Running coupling corrections to 
inclusive gluon production}

\author{W. A. Horowitz,$^{1, 2}$ \ Yuri V. Kovchegov$^{\ 1}$}

\address{$^1$Department of Physics, The Ohio State University, Columbus,
OH 43210, USA \\

$^2$Department of Physics, University of Cape Town, Rondebosch
7701, South Africa} 

\ead{wa.horowitz@uct.ac.za, yuri@physics.ohio-state.edu}

\begin{abstract}
  We calculate running coupling corrections for the lowest-order gluon
  production cross section in high energy hadronic and nuclear
  scattering using the BLM scale-setting prescription. At leading
  order there are three powers of fixed coupling; in our final answer,
  these three couplings are replaced by seven factors of running
  coupling: five in the numerator and two in the denominator, forming
  a `septumvirate' of running couplings, analogous to the
  `triumvirate' of running couplings found earlier for the small-$x$
  BFKL/BK/JIMWLK evolution equations. It is interesting to note that
  the two running couplings in the denominator of the `septumvirate'
  run with complex-valued momentum scales, which are complex
  conjugates of each other, such that the production cross section is
  indeed real. We use our lowest-order result to conjecture how
  running coupling corrections may enter the full fixed-coupling
  $k_T$-factorization formula for gluon production which includes
  non-linear small-$x$ evolution.
\end{abstract}


\section{Introduction}

This proceedings contribution is based on \cite{Horowitz:2010yg}.

While an exact analytic formula for gluon production in $AA$
collisions in the Color Glass Condensate (CGC) framework is still not
known, we do know that in $pp$ and $pA$ collisions gluon production at
the level of classical gluon fields and leading-$\ln 1/x$ nonlinear
quantum evolution is given by the $k_T$-factorization formula
\cite{Kovchegov:2001sc,Braun:2000bh}:
\begin{equation}\label{ktfact}
  \frac{d \sigma}{d^2 k_T \, dy} \, = \, \frac{2 \, \as}{C_F} \,
  \frac{1}{{\bm k}^2} \, \int d^2 q \, \phi_p ({\bm q}, y) \, \phi_{A}
  ({{\bm k} - \bm q}, Y-y).
\end{equation}
Here $Y$ is the total rapidity interval of the collision, $C_F =
(N_c^2 -1)/2 N_c$, boldface variables denote two-component transverse
plane vectors ${\bm k} = (k^1 , k^2)$, and $\phi_p$, $\phi_A$ are the
unintegrated gluon distributions in the proton and the nucleus,
respectively, which are defined by \cite{Kovchegov:2001sc}
\begin{equation}\label{ktglueA}
  \phi_A ({\bm k}, y) \, = \, \frac{C_F}{\as \, (2 \pi)^3} \, \int d^2 b \, 
d^2 r \, e^{- i {\bm k} \cdot {\bm r}} \ \nabla^2_r \, N_G ({\bm r},
{\bm b}, y)
\end{equation}
and 
\begin{equation}\label{ktgluep}
\phi_p ({\bm k}, y) \, = \, \frac{C_F}{\as \, (2 \pi)^3} \, \int d^2 b \, 
d^2 r \, e^{- i {\bm k} \cdot {\bm r}} \ \nabla^2_r \, n_G ({\bm r},
{\bm b}, y).
\end{equation}
In \eq{ktglueA} the quantity $N_G ({\bm r}, {\bm b}, y)$ denotes the
forward scattering amplitude for a gluon dipole of transverse size
$\bm r$ with its center located at the impact parameter $\bm b$
scattering on a target nucleus with total rapidity interval $y$.
$N_G ({\bm r}, {\bm b}, y)$ can, in general, be found from the JIMWLK
evolution equation. In the large-$N_c$ limit it is related to the
quark dipole forward scattering amplitude on the same nucleus $N ({\bm
  r}, {\bm b}, y)$ by
\begin{equation}\label{2NN}
  N_G ({\bm r}, {\bm b}, y) \, = \, 2 \, N ({\bm r}, {\bm b}, y) - N
  ({\bm r}, {\bm b}, y)^2,
\end{equation}
where $N ({\bm r}, {\bm b}, y)$ can be found from the BK evolution
equation.  The quantity $n_G ({\bm r}, {\bm b}, y)$ from \eq{ktgluep}
is also a gluon dipole amplitude, but taken in a dilute regime, where
it is found by solving the linear Balitsky-Fadin-Kuraev-Lipatov (BFKL)
evolution equation.

\eq{ktfact} for the gluon production was derived in the fixed coupling
approximation. However, the dipole amplitudes $N ({\bm r}, {\bm b},
y)$, $N_G ({\bm r}, {\bm b}, y)$ and $n_G ({\bm r}, {\bm b}, y)$ are
now known for the running coupling case due to the completion of the
running coupling calculations for the BFKL/BK/JIMWLK evolution
equations \cite{Kovchegov:2006vj,Balitsky:2006wa}.  Using the
running-coupling BK (rcBK) equation, the calculational framework
presented above, when applied to heavy ion collisions by replacing
$\phi_p \rightarrow \phi_A$ in \eq{ktfact}, and implemented with a
careful inclusion of the nuclear geometry fluctuations, led to the
prediction made in \cite{ALbacete:2010ad} of the charged particle
multiplicity as a function of the collision centrality for the LHC.
This prediction was confirmed by the ALICE data in
\cite{Aamodt:2010cz}.  However, at the time the prediction
\cite{ALbacete:2010ad} was made, the scales of the couplings
explicitly shown in Eqs. \peq{ktfact}, \peq{ktglueA}, and
\peq{ktgluep} were not known and had to be modeled. Our goal here is
to fix the scales of those couplings for the lowest-order gluon
production.


\section{Running coupling corrections: strategy}

To include running coupling corrections we follow the BLM
scale-setting procedure \cite{BLM}, which is known to be correct at
least at the leading order in $1/N_c^2$. One first needs to resum the
contribution of all quark bubble corrections giving powers of $\amu \,
N_f$, with $N_f$ the number of quark flavors and $\amu$ the physical
coupling at some arbitrary renormalization scale $\mu$. We then
complete $N_f$ to the full beta-function by replacing
\begin{equation}\label{repl}
N_f \rightarrow - 6 \, \pi \, \beta_2
\end{equation} 
in the obtained expression. Here
\begin{equation}\label{beta}
\beta_2 = \frac{11 N_c - 2 N_f}{12 \, \pi}  
\end{equation} 
is the one-loop QCD beta-function. After this, the powers of $\amu \,
\beta_2$ should combine into physical running couplings 
\begin{equation}\label{as_geom}
  \as (Q^2) \, = \, \frac{\amu}{1 + \amu \, \beta_2 \, \ln
    \frac{Q^2}{\mu^2}}
\end{equation}
at various momentum scales $Q$ which would follow from this
calculation. We use the $\overline {\mbox MS}$ renormalization scheme.

As was originally argued in \cite{Kovchegov:2007vf}, including running
coupling corrections into the diagrams of \fig{fig1} below assuming
that only a gluon can be produced in the final state would leave one
factor of the coupling at an arbitrary renormalization scale $\mu$.
Following \cite{Kovchegov:2007vf} we rectify the problem by redefining
the gluon production cross-section to include production of collinear
gluon--gluon and quark--anti-quark pairs with the invariant mass lower
than some collinear IR cutoff $\Lambda_{coll}^2$. This new observable
is completely $\mu$-independent and expressible in terms of the
running coupling constants.


\section{Running coupling corrections to LO gluon production}

\begin{figure}[th]
\begin{center}
\leavevmode
\includegraphics[width=15cm]{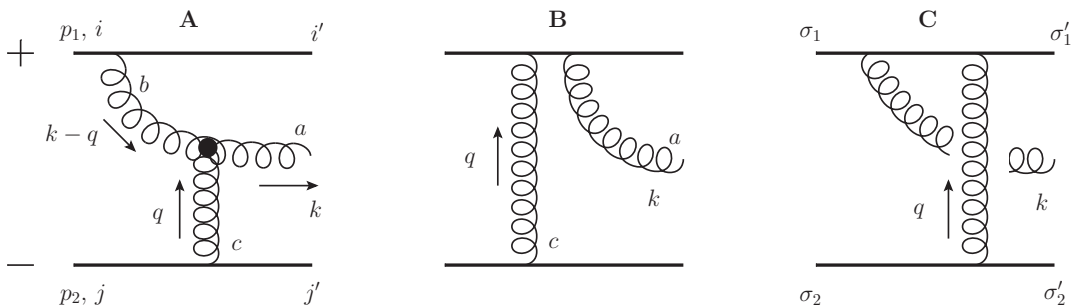}
\end{center}
\caption{Diagrams contributing to the lowest-order gluon production in 
  quark-quark scattering at high energy in the $A^+ =0$ light-cone
  gauge. }
\label{fig1}
\end{figure}

Gluon production at the lowest order in the coupling is shown in
\fig{fig1} in the $A^+ =0$ light-cone gauge. At this order the
unintegrated gluon distribution is
\begin{equation}\label{phi_lo}
  \phi ({\bm k}, y) \, = \, \frac{\as \, C_F}{\pi} \, \frac{1}{{\bm k}^2}.
\end{equation}
such that \eq{ktfact} reduces to 
\begin{equation}\label{lip2}
  \frac{d \sigma}{d^2 k_T \, dy} \, = \, \frac{2 \, \as^3 \,
    C_F}{\pi^2} \, \frac{1}{{\bm k}^2} \, \int \, \frac{d^2 q}{{\bm
      q}^2 \, ({\bm k} - {\bm q})^2}. 
\end{equation}
Our goal is to set the scales for the three couplings in \eq{lip2}.

The final result for the lowest-order gluon production cross section
with the running coupling corrections included is \cite{Horowitz:2010yg}
\begin{equation}\label{rc_incl}
  \frac{d \sigma}{d^2 k_T \, dy} \, = \, \frac{2 \, C_F}{\pi^2} \,
  \frac{ \as \left( \Lambda_{coll}^2 \right)}{{\bm k}^2} \, \int \frac{d^2 q}{{\bm q}^2 \, ({\bm k} - {\bm q})^2} \, \frac{\as^2 \left( {\bm q}^2 \right) \, \as^2 \left( ({\bm k} - {\bm q})^2 \right)}{\as \left( Q^2 \right) \, \as \left( Q^{* \, 2} \right)}
\end{equation}
with the momentum scale $Q$ being a complicated function of ${\bm q}$
and ${\bm k} - {\bm q}$ given in \cite{Horowitz:2010yg}. 
An interesting feature is that the scale $Q^2$ is complex-valued! The
cross-section \peq{rc_incl} is, of course, real, as it contains a
complex-valued coupling constant multiplied by its conjugate, $\as
\left( Q^2 \right) \, \as \left( Q^{* \, 2} \right)$. \eq{rc_incl}
clearly looks like the fixed-coupling cross-section \peq{lip2} with
three factors of fixed-coupling replaced by the seven running
couplings: we chose to refer to this structure as {\sl the
  septumvirate} of couplings \cite{Horowitz:2010yg}.

An important feature of the scale $Q$ is that in the $\bm q
\rightarrow 0$ (or $\bm q \rightarrow \bm k$) limit
\begin{equation}
  \ln \frac{Q^2}{\msbar^2} \, = \, \ln \frac{{\bm k}^2}{\msbar^2} +
  \frac{1}{2}
\end{equation}
reducing \eq{rc_incl} to 
\begin{equation}\label{rc_incl_approx2}
  \frac{d \sigma}{d^2 k_T \, dy} \, \approx \, \frac{4 \, C_F}{\pi} \,
  \frac{\as \left( \Lambda_{coll}^2 \right) \, \as ({\bm k}^2) \,
    \as (Q_s^2)}{\left( {\bm k}^2 \right)^2} \, \ln \frac{{\bm
      k}^2}{Q_s^2}.
\end{equation}


\section{Ansatz for the running coupling corrections in the $k_T$-factorization formula}

Using the lowest-order expression \peq{rc_incl} we would like to
conjecture the following running-coupling generalization of
\eq{ktfact}:
\begin{equation}\label{rc_fact}
  \frac{d \sigma}{d^2 k_T \, dy} \, = \, \frac{2 \, C_F}{\pi^2} \,
  \frac{1}{{\bm k}^2} \int d^2 q \ {\overline \phi}_p ({\bm q}, y)
  \, {\overline \phi}_A ({\bm k} - {\bm q}, Y-y) \, \frac{\as \left(
      \Lambda_{coll}^2 \right)}{\as \left( Q^2 \right) 
  \, \as \left( Q^{* \, 2} \right)}
\end{equation}
with the new (rescaled) distribution functions defined by (cf. Eqs.
\peq{ktglueA} and \peq{ktgluep})
\begin{equation}\label{rc_ktglueA}
  {\overline \phi}_A ({\bm k}, y) \, = \, \frac{C_F}{(2 \pi)^3} \,
  \int d^2 b \, d^2 r \, e^{- i {\bm k} \cdot {\bm r}} \ \nabla^2_r \,
  N_G ({\bm r}, {\bm b}, y)
\end{equation}
and 
\begin{equation}\label{rc_ktgluep}
  {\overline \phi}_p ({\bm k}, y) \, = \, \frac{C_F}{(2 \pi)^3} \,
  \int d^2 b \, d^2 r \, e^{- i {\bm k} \cdot {\bm r}} \ \nabla^2_r \,
  n_G ({\bm r}, {\bm b}, y).
\end{equation}
One may use this ansatz, along with $N_G$ obtained from the rcBK
evolution to further improve the existing CGC phenomenology for
particle multiplicity along with its centrality and rapidity
dependence at RHIC and LHC.


\section*{Acknowledgments}

This research is sponsored in part by the U.S. Department of Energy
under Grant No. DE-SC0004286.


\section*{References}



\end{document}